\documentclass{ptptex}
\usepackage{epsfig}
\usepackage{graphicx}
\usepackage{amssymb}
\def\be{\begin{equation}}
\def\ee{\end{equation}}
\def\ba{\begin{array}}
\def\ea{\end{array}}
\def\beqn{\begin{eqnarray}}
\def\eeqn{\end{eqnarray}}

\def\bt{\begin{tabular}}
\def\et{\end{tabular}}
\def\bc{\begin{center}}
\def\ec{\end{center}}
\notypesetlogo 

\title{Investigating texture six zero lepton mass matrices}

\author{Neelu \textsc{Mahajan}$^1$, Monika \textsc{Randhawa}$^2$, 
Manmohan \textsc{Gupta}$^3$, P.S. \textsc{Gill}$^4$}

\inst{$^1$ \it D. A. V. College, Chandigarh, India.\\
$^2$  University Institute of Engineering and Technology,
Panjab University, Chandigarh, India.\\
$^3$  Department of
Physics, Centre of Advanced Study, Panjab University, Chandigarh, India.\\
 $^4$ Sri Guru Gobind Singh College,
Chandigarh, India.}
\recdate{}
\abst{
 Texture six zero Fritzsch like as well as non Fritzsch like
 hermitian lepton mass matrices (144 combinations in all) have been
investigated for both Majorana and Dirac neutrinos for their
compatibility with the current neutrino oscillation data, keeping
in mind the hierarchy of neutrino masses. All the combinations
considered here for Majorana neutrino masses are ruled out by the
existing data in the case of inverted hierarchy and degenerate
scenario. For Majorana neutrinos with normal hierarchy, only 16
combinations can accommodate the experimental data. Assuming
neutrinos to be Dirac particles, normal hierarchy, inverted
hierarchy as well as degenerate neutrinos are ruled out for all
combinations of texture 6 zero hermitian mass matrices.}
\begin{document}
\maketitle
\section{Introduction}The large value of $\theta_{13}$ determined by the reactor  neutrino experiments \cite{reac} has not only thrown open the door for search of CP violation in the lepton sector, but has also provided impetus to formulate the theories for understanding the origin of neutrino masses and their mixing.
Apart from the recent $\theta_{13}$ measurement, there also has been
considerable progress in the measurement of neutrino mass square
differences and mixing angles $\theta_{12}$ and $\theta_{23}$ in the last few  years. However, not much information is available about the leptonic CP violation phase $\delta_l$ from the neutrino oscillation data. Also, the absolute neutrino mass scale is still unknown. Further, the presently available neutrino oscillation
data does not throw any light on the neutrino mass hierarchy,
which may
 be normal/inverted hierarchy and may even be degenerate.
Furthermore, the situation becomes complicated when one realizes
that neutrino masses are much smaller than charged fermion masses
as well as it is yet not clear whether neutrinos are Dirac or
Majorana particles. 

In the absence of a convincing fermion flavor
theory, several approaches have been considered to understand
fermion mass generation problem, e.g., texture zeros,\cite{text} \ 
seesaw mechanism,\cite{seesaw} \ radiative mechanisms,\cite{rad} \ 
flavor symmetries,\cite{flasym} \ extra dimensions,\cite{extra} \
etc.. In this context, texture specific mass matrices have
received a good deal of attention in the literature, in
particular,
 Fritzsch-like texture specific mass matrices seem to be very
helpful in understanding the pattern of quark mixings and CP
violation.\cite{fri2000}\tocite{neelu} \ Taking clues from the
success of these  matrices in the context of quarks, several
attempts have been made to consider texture specific lepton mass
matrices \cite{fri2004}\tocite{nonflavor} for explaining the
pattern of neutrino masses and mixings. In the absence of a
sufficient amount of data regarding neutrino masses and mixings,
it would require a very careful scrutiny of all possible textures
to find viable structures, which are compatible with data and
theoretical ideas, so that these are kept in mind while
formulating mass matrices at the grand unified theory (GUT) scale.

In the quark sector, both Fritzsch like as well as non Fritzsch
like texture six zero mass matrices have been completely ruled out.\cite{neelu}
 \  In leptonic sector, most of the analyses
 have been carried out in the flavor basis.\cite{flavor} \  
In non flavor basis, where both charged lepton mass matrix $M_{l}$
and neutrino mass matrix $M_{\nu}$ are three zero type, the number
of possibilities for texture six zero mass matrices becomes very
large. These possibilities have been explored in the literature
\cite{nfXing} both for Majorana as well as for Dirac neutrinos,
however adequate attention has not been given to the cases for
inverted hierarchy and degenerate scenarios. Also, it is desirable
to note that Dirac neutrinos have not yet been ruled out by the
experiments,\cite{dirac} \ therefore it becomes interesting in the
case of texture six zero mass matrices to carry out a detailed
comparison for Dirac like as well as Majorana like neutrinos in
the case of normal, inverted and degenerate cases. This exercise
becomes all the more interesting in view of the refinements of
data and advocacy of quark lepton symmetry.\cite{qlu}

The purpose of the present paper is to update the analysis of Zhou and Xing 
\cite{nfXing} for all possibilities of texture
six zero lepton mass matrices
as well as to extend this analysis to the
case of inverted and degenerate neutrino masses.
To preserve the parallelism between quarks and leptons
only those neutrino mass matrices have been considered, which are
consistent with the requirement of non zero and distinct
neutrino masses.
 Following our analysis in the quark sector,\cite{neelu} \
the mass matrix for leptons and neutrinos are taken to be
hermitian. For the sake of completion, we have also investigated
the cases corresponding to charged leptons being in the flavor
basis. It would also be a desirable exercise to calculate
phenomenological quantities, such as  effective neutrino mass
$\langle m_{ee} \rangle $ related to
neutrinoless double beta decay, Jarlskog's rephasing
invariant parameter in the leptonic sector $J_l$ and the
corresponding  Dirac like CP
 violating phase $\delta_l$ for the viable cases.

The detailed plan of the paper is as follows. In Section (2), we
present the essentials of the formalism connecting the mass matrix
to the neutrino mixing matrix. Inputs used in the present analysis
have been given in Section (3). In Section (4), various combinations
of texture six zero mass matrices have been given. In Sections (5)
and (6), for Majorana and Dirac neutrinos respectively, the detailed
calculations pertaining  to normal, degenerate and inverted
hierarchy have been discussed. Finally, Section (7), summarizes our
conclusions.

\section{Construction of PMNS matrix from mass matrices\label{form}}

To begin with, we present the Fritzsch like hermitian texture six zero lepton
 mass matrices, e.g.,
 \be
 M_{l}=\left( \ba{ccc}
0 & A _{l} & 0      \\ A_{l}^{*} & 0 &  B_{l}     \\
 0 &     B_{l}^{*}  &  C_{l} \ea \right), \qquad
M_{\nu D }=\left( \ba{ccc}
0 & A _{\nu D} & 0      \\ A_{\nu D}^{*} & 0  &  B_{\nu D}     \\
 0 &     B_{\nu D}^{*}  &  C_{\nu D} \ea \right), \qquad
\label{frzmm5}
 \ee
$M_l$ and $M_{\nu D}$ respectively corresponding to Dirac-like charged
 lepton and neutrino mass matrices. It may be noted that each of the
 above matrix is texture three zero type with $A_{l(\nu D)}
=|A_{l(\nu D)}|e^{i\alpha_{l(\nu D)}}$
 and $B_{l(\nu D)} = |B_{l(\nu D)}|e^{i\beta_{l(\nu D)}}$.
For Majorana neutrinos, the neutrino mass matrix
$M_{\nu}$ is given by seesaw mechanism,\cite{seesaw} \ for example,
\be M_{\nu}=-M_{\nu D}^T\,(M_R)^{-1}\,M_{\nu D},
\label{seesaweq} \ee \noindent
 where $M_{\nu D}$ and $M_R$ are
respectively, the Dirac neutrino mass matrix and the right-handed
Majorana mass matrix.
It may be mentioned that for both
Majorana as well as Dirac neutrinos the texture is imposed only on
$M_{\nu D}$, with no such restriction on $M_{\nu}$ for Majorana case.
In the absence of any guidelines for the right handed Majorana mass matrix
 $M_{R}$, to keep the number of parameters under control, it would be
desirable to keep its structure as simple as possible. Therefore, 
following Fukugita {\it et al},\cite{yana} \  we
take $M_R = m_R I$,    where $I$ is the unity matrix and $m_R$ denotes a
very large mass scale. 
Here, it is pertinent to mention that any mechanism leading to texture zeros in  
 $M_{\nu D}$ does not  necessarily require a diagonal
form of  $M_R$, therefore one may as well consider a more general form of $M_R$ with or without texture zeros. However, since the seesaw framework contains more free 
parameters than can be obtained from the low energy data, therefore considering a texture zero model for $M_R$  leads to reduction in the number of parameters, and thus
enhances the predictive power of the model. In this regard, recently Fritzsch et al \cite{frzzhou}  have considered a model wherein a  three zero texture structure is imposed on charged lepton mass matrix $M_l$, the Dirac neutrino mass matrix $M_{\nu D}$ and also on heavy right-handed Majorana mass matrix $M_R$. Therefore,  to widen the scope of the paper as well as for completeness sake,  in the Appendix (B), we have briefly discussed the cases where we consider a parallel texture three zero structure 
for $M_R$ and $M_{\nu D}$.

To fix the notations and conventions as well as to facilitate the
 understanding of inverted hierarchy case and its relationship to
 normal hierarchy case, we detail the formalism connecting the mass
 matrices to the neutrino mixing matrix. To facilitate the
diagonalization of $M_k$, where $k=l,\nu D$, the complex mass matrix
 $M_k$ can be expressed as
\be
M_k= Q_k M_k^r P_k \,  \label{mk} \ee or  \be M_k^r= Q_k^{\dagger}
M_k P_k^{\dagger}\,, \label{mkr} \ee where $M_k^r$ is a real
symmetric matrix with real eigenvalues and $Q_k$ and $P_k$ are
diagonal phase matrices. For the hermitian mass matrix $Q_k=
P_k^{\dagger}$. In general, the real matrix $M_k^r$ is
diagonalized by the orthogonal transformation $O_k$, e.g., \be
M_k^{diag}= {O_k}^T M_k^r O_k \,, \label{mkdiag} \ee which on
using Equation (\ref{mkr}) can be rewritten as \be M_k^{diag}=
{O_k}^T Q_k^{\dagger} M_k P_k^{\dagger} O_k \,. \label{mkdiag2}
\ee To facilitate the construction of diagonalization
transformations for different hierarchies, we introduce a diagonal
phase matrix $\xi_k$ defined as $ {\rm diag} (1,\,e^{i \pi},\,1)$
for the case of normal hierarchy and as $ {\rm diag} (1,\,e^{i
\pi},\,e^{i \pi})$ for the case of inverted hierarchy. Equation
(\ref{mkdiag2}) can now be written as \be \xi_k M_k^{diag}=
{O_k}^T Q_k^{\dagger} M_k P_k^{\dagger} O_k \,, \label{mkdiag3}
\ee which can also be expressed as \be M_k^{diag}= \xi_k^{\dagger}
{O_k}^T Q_k^{\dagger} M_k P_k^{\dagger} O_k \,. \label{mkdiag4}
\ee Making use of the fact that $O_k^*=O_k$ it can be further
expressed as
\be
M_k^{diag}=(Q_k O_k \xi_k)^{\dagger} M_k (P_k^{\dagger}
O_k),\label{mkeq} \ee from which one gets \be M_k=Q_k O_k \xi_k
M_k^{diag} O_k^T P_k.\label{mkeq2} \ee

The case of leptons is fairly straight forward, whereas in the
case of neutrinos, the diagonalizing transformation is hierarchy
specific as well as requires some fine tuning of the phases of the
right handed mass matrix $M_R$. To clarify this point
further, in analogy with Equation (\ref{mkeq2}), we can express
$M_{\nu D}$ as \be M_{\nu D}=Q_{\nu D} O_{\nu D} \xi_{\nu D}
M_{\nu D}^{diag} O_{\nu D}^T P_{\nu D}.\label{mnud} \ee
Substituting the above value of $M_{\nu D}$ in Equation
(\ref{seesaweq}) one obtains
\be
M_{\nu}=-(Q_{\nu D} O_{\nu D} \xi_{\nu D} M_{\nu D}^{diag} O_{\nu
D}^T P_{\nu D})^T (M_R)^{-1} (Q_{\nu D} O_{\nu D} \xi_{\nu D}
M_{\nu D}^{diag} O_{\nu D}^T P_{\nu D}), \ee which, on using
$P_{\nu D}^T = P_{\nu D}$, $Q_{\nu D}^T = Q_{\nu D}$, 
can further be written as
\be
M_{\nu}=-P_{\nu D} O_{\nu D} M_{\nu D}^{diag} \xi_{\nu D} O_{\nu
D}^T Q_{\nu D} (M_R)^{-1} Q_{\nu D} O_{\nu D} \xi_{\nu D} M_{\nu
D}^{diag} O_{\nu D}^T P_{\nu D}, \ee wherein, assuming fine
tuning, the phase matrices $Q_{\nu D}^T$ and $Q_{\nu D}$ along
with $-M_R$ can be taken as $m_R ~{\rm diag} (1,1,1)$ as well as
using the unitarity of $\xi_{\nu D}$ and orthogonality of $O_{\nu
D}$, the above equation can be expressed as
\be
M_{\nu}= P_{\nu D} O_{\nu D} \frac{(M_{\nu
D}^{diag})^2}{m_R} O_{\nu D}^T P_{\nu D}. \label{mnu} \ee

The lepton mixing matrix or the Pontecorvo-Maki-Nakagawa-Sakata
(PMNS) matrix \cite{PMNS} $U$
can be obtained from the matrices used
for diagonalizing the mass matrices $M_l$ and $M_{\nu}$ and is
expressed as
 \be
U =(Q_l O_l \xi_l)^{\dagger} (P_{\nu D} O_{\nu D}). \label{mix}
\ee Eliminating the phase matrix $\xi_l$ by redefinition of the
charged lepton phases, the above equation becomes
\be
 U = O_l^{\dagger} Q_l P_{\nu D} O_{\nu D} \,, \label{mixreal} \ee
where $Q_l P_{\nu D}$, without loss of generality, can be taken as
$(e^{i\phi_1},\,e^{i\phi_2},\,e^{i\phi_3})$;
$\phi_1$, $\phi_2$ and $\phi_3$ being
related to the phases of mass matrices and can be treated as free
parameters.

\section{Inputs used in the present analysis\label{in}}
Before going into the details of the analysis, we would like to
mention some of the essentials pertaining to various inputs. The
inputs for masses and mixing angles used in the present analysis
at 3$\sigma$ C.L. are as follows,\cite{foglinew}
\be
 \Delta m_{12}^{2} = (6.99 - 8.18)\times
 10^{-5}~\rm{eV}^{2},~~ \Delta {\it m}_{23}^{2}=\left\{
\begin{array}{cc}(2.19 - 2.62)\times 10^{-3}~ \rm{eV}^{2} & NH \\
 (2.17 - 2.61)\times 10^{-3}~ \rm{eV}^{2}& IH
\end{array} \right.
 \label{solatmmass} \ee
\be
{\rm sin}^2\,\theta_{12}  =  0.259 - 0.359, ~~
 {\rm sin}^2\,\theta_{23}  = \left\{
\begin{array}{cc} 0.331 - 0.637 & NH \\ 0.335 - 0.663 & IH
\end{array} \right.,
 {\rm sin}^2\,\theta_{13} =\left\{
\begin{array}{cc} 0.0169 - 0.0313 & NH \\ 0.0171 - 0.0315 & IH 
\end{array}
\right.. \label{s13} \ee \\
where NH and IH corresponds to normal
hierarchy and inverted hierarchy respectively.

For the purpose of calculations, we have taken the lightest
neutrino mass and the phases $\phi_1$, $\phi_2$ and $\phi_3$ as free
parameters. The other two masses are constrained by $\Delta
m_{12}^2 = m_{\nu_2}^2 - m_{\nu_1}^2 $ and $\Delta m_{23}^2 =
m_{\nu_3}^2 - m_{\nu_2}^2 $ in the normal hierarchy case defined
as $m_{\nu_1}<m_{\nu_2}\ll m_{\nu_3}$ and also valid for the
degenerate case defined as $m_{\nu_1} \lesssim m_{\nu_2} \sim
m_{\nu_3}$ and by $\Delta m_{23}^2 = m_{\nu_2}^2 - m_{\nu_3}^2$ in
the inverted hierarchy case defined as $m_{\nu_3} \ll m_{\nu_1} <
m_{\nu_2}$. It may be noted that lightest neutrino mass
corresponds to $m_{\nu_1}$ for the normal hierarchy case and to
$m_{\nu_3}$ for the inverted hierarchy case. The explored range of
lightest neutrino mass is taken to be
$0.0001\,\rm{eV}-1.0\,\rm{eV}$ as our results remain unaffected
even if the range is extended further. In the absence of any
constraints on the phases $\phi_1$, $\phi_2$
 and  $\phi_3$,
 these have been given full variation from 0 to $2\pi$.

\section{Texture six zero lepton mass matrices}

\begin{table}
\caption{Table showing
various patterns of texture three zero mass matrices classified
into two classes, Class I and II.}\label{t12}
\bt{ccc} \hline\hline
  & Class I  & Class II \\ \hline
$a$ & $\left ( \ba{ccc} {\bf 0} & Ae^{i\alpha} & {\bf 0} \\
Ae^{-i\alpha}  & {\bf 0} & Be^{i\beta} \\ {\bf 0} & Be^{-i\beta}
& C \ea \right )$  & $\left ( \ba{ccc} {\bf 0} & Ae^{i\alpha} &
{\bf 0} \\ Ae^{-i\alpha}  & B & {\bf 0} \\ {\bf 0} & {\bf 0}  & C
\ea \right )$ \\ & & \\ $b$ &  $\left ( \ba{ccc} {\bf 0} &{\bf 0} &
Ae^{i\alpha} \\ {\bf 0}  & C & Be^{i\beta} \\Ae^{-i\alpha} &
B^{-i\beta}  & {\bf 0} \ea \right )$  &
 $\left ( \ba{ccc} {\bf 0} & {\bf 0} & Ae^{i\alpha}
 \\ {\bf 0}  & C & {\bf 0} \\ Ae^{-i\alpha}  & {\bf 0}  &
B \ea \right )$ \\ & & \\ $c$ &  $\left ( \ba{ccc} {\bf 0} &
Ae^{i\alpha} & Be^{i\beta} \\ Ae^{-i\alpha}  & {\bf 0} & {\bf 0}
\\ Be^{-i\beta} & {\bf 0}  & C \ea \right )$  &
 $\left ( \ba{ccc} B & Ae^{i\alpha} &
{\bf 0} \\ Ae^{-i\alpha}  & {\bf 0} & {\bf 0} \\ {\bf 0} & {\bf 0}
& C \ea \right )$ \\  & & \\ $d$ &  $\left ( \ba{ccc} C &
Be^{i\beta} & {\bf 0}
\\ Be^{-i\beta}  & {\bf 0} & Ae^{i\alpha}\\ {\bf 0}  & Ae^{-i\alpha} &
{\bf 0} \ea \right )$  &
 $\left ( \ba{ccc} C & {\bf 0} & {\bf 0}
 \\ {\bf 0}  & B & Ae^{i\alpha} \\ {\bf 0} & Ae^{-i\alpha}  &
{\bf 0} \ea \right )$ \\ & & \\ $e$ &  $\left ( \ba{ccc} {\bf 0} &
Be^{i\beta}  & Ae^{i\alpha} \\ Be^{-i\beta}  & C & {\bf 0} \\
Ae^{-i\alpha}  & {\bf 0}  &
 {\bf 0} \ea \right )$  &
 $\left ( \ba{ccc} B & {\bf 0} & Ae^{i\alpha}
\\ {\bf 0} & C &  {\bf 0} \\ Ae^{-i\alpha}  & {\bf 0}  &
{\bf 0}  \ea \right )$ \\  & & \\ $f$ & $\left ( \ba{ccc} C & {\bf
0} & Be^{i\beta}
 \\ {\bf 0}  & {\bf 0}  & Ae^{i\alpha} \\Be^{-i\beta} & Ae^{-i\alpha}  &
{\bf 0} \ea \right )$  &
 $\left ( \ba{ccc} C & {\bf 0} &{\bf 0}
 \\ {\bf 0}  & {\bf 0} & Ae^{i\alpha} \\ {\bf 0} & Ae^{-i\alpha}  &
B \ea \right )$ \\  & & \\  \hline \et 
\end{table}

To begin with, we enumerate the number of possibilities for the
texture six zero lepton mass matrices. It is easy to see from
Equation (\ref{frzmm5}) that  there are 20 possible patterns of
texture three zero hermitian mass matrices, which differ from each
other with regard to the position of zeros in the structure of
mass matrix.

 Texture six zero mass matrices
are obtained when both $M_l$ and $M_{\nu D}$  are texture three zero type,
implying that  there will be 400
combinations of texture six zero lepton mass matrices. As mentioned
earlier, in the present work, we have considered only those mass
matrices which lead to non zero and distinct mass eigen values,
therefore imposing the trace and determinant condition on the mass
matrix. i.e. Det $M_{l,\nu D}  \neq 0$ and Trace $M_{l,\nu D}  \neq 0$,
we are left with 12 patterns, classified into 2 distinct classes,
 depending upon the diagonalization equations
these satisfy as given in Table (\ref{t12}). Details of the
diagonalization equations for these 12 mass matrices can be looked
up in our earlier work.\cite{neelu} \  Matrices $M_{l}$ and $M_{\nu
D}$ each can correspond to any of the 12 patterns, therefore
yielding 144 possible combinations of texture six zero lepton mass
matrices which in principle can yield neutrino mixing matrix.
These 144 combinations form 4 different categories, e.g.;\\ \\
Category 1: $M_{l}$ from Class I and $M_{\nu D}$ from Class I. \\
Category 2: $M_{l}$ from Class II and $M_{\nu D}$ from Class II.
\\ Category 3: $M_{l}$ from Class I and $M_{\nu D}$ from Class II.
\\ Category 4: $M_{l}$ from Class II and $M_{\nu D}$ from Class I.
\\

Each category corresponds to 36 combinations which need exhaustive
analysis. For all these combinations, we have considered  the
cases of normally hierarchical,  inversely hierarchical and
degenerate Majorana  as well as  Dirac neutrinos.

\section{Majorana Neutrinos}

We have analyzed 144 combinations
of texture six zero lepton mass matrices corresponding to
 normal hierarchy of Majorana neutrinos,
by confronting their corresponding mixing matrix against
latest neutrino oscillation data given in Section (\ref{in}).
 It may be mentioned again that for
Majorana neutrinos, the texture three zero structure is imposed on
$M_{\nu D}$ as given in Equation (\ref{seesaweq}).

In literature, the details of Fritzsch like texture six zero
lepton mass matrices \cite{fri2004}\tocite{nonflavor}
 have been presented by several
authors, however, for non-Fritzsch like mass matrices similar
details have not been presented.
 To detail the
methodology, we consider a combination corresponding to
Category (4), where both $M_{l}$ and $M_{\nu D}$ are non Fritzsch
type and as mentioned earlier $M_R$ is diagonal. For example;
\be
 M_{l}=\left( \ba{ccc}

C_{l} & 0 & 0      \\ 0 & D_{l} & A_{l}     \\
 0 & A_{l}^{*}   & 0 \ea \right),~~~
M_{\nu D}=\left( \ba{ccc} 0 & 0 & A_{\nu D}        \\ 0 &
C_{\nu D} &  B_{\nu D}     \\
A_{\nu D}^{*} &  B_{\nu D}^{*}  &  0 \ea \right),~~~ M_R=m_R 
\left( \ba{ccc} 1 & 0 & 0   \\ 0 & 1 &  0  \\
0 &  0  & 1 \ea \right). \label{nfe}
 \ee

The diagonalizing transformations for these matrices can easily be obtained in terms of neutrino masses $m_{\nu_1}$, $m_{\nu_2}$ and $m_{\nu_3}$ for both normal and inverted hierarchy and are given in Appendix (A). The phase
matrix  $ Q_{l}P_{\nu D}$ for this particular combination is given
as $(e^{i\phi_1},\,e^{i\phi_2},\,1).$ The scanned ranges of
lightest neutrino mass $m_{\nu_1}$ and phases $\phi_1$ and
$\phi_2$ have been given in Section (\ref{in}) and
 $\Delta m_{12}^2$, $ \Delta m_{23}^2$ and
the mixing angles have been constrained, as given by Equations
(\ref{solatmmass}) and
 (\ref{s13}).

Using  Equation (\ref{mixreal}), the PMNS matrix $U$,
obtained for this particular combination is
 \be U = \left( \ba{ccc}
0.78-0.84 & ~0.52-0.61 & ~0.12-0.16     \\ 0.39-0.49 & ~0.40-0.53
& ~0.72-0.79     \\ 0.29-0.43 & ~0.67-0.72   & ~0.59-0.68 \ea
\right),\label{uobt}\ee in good agreement with the ranges of
mixing matrix element given
 by Garcia et al.\cite{garcia} \ at 3$\sigma$ C.L.

\begin{figure}[hbt]
\centerline{\includegraphics[width=15cm,height=12cm]{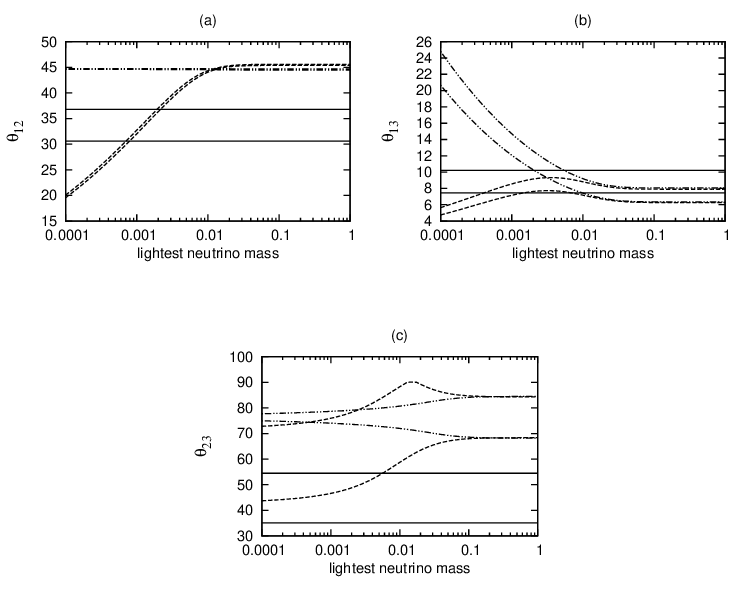}}
 \caption{Plots showing variation of mixing angles $\theta_{12}$, $\theta_{13}$
and $\theta_{23}$ with lightest neutrino mass for $\rm II_d \rm
I_b $ combination given in Equation (\ref{nfe}) for Majorana
neutrinos. The dashed lines and the dot-dashed lines depict the
limits obtained assuming normal and inverted hierarchy
respectively, the solid horizontal lines show the experimental
3$\sigma$ limits.}
 \label{mni}
\end{figure}

  To graphically show the
viability of this particular combination of texture six zero mass
matrices, in Figure (1) we have plotted the lightest neutrino mass
against the mixing angle $\theta_{12}$, $\theta_{13}$ and
$\theta_{23}$ by giving full variation to input parameters. The
dashed lines depict the limits obtained by assuming normal
hierarchy and the solid horizontal lines show the experimental
 3$\sigma$ limits of the plotted mixing angles. A general
look at the Figure (1), shows that mixing
angles are well within their experimental ranges for a common neutrino
mass range for normal hierarchy. Moreover, one can see that
the neutrino masses are following strict normal hierarchy.

 In a similar manner, one can check the viability of the above
set of mass matrices for inversely hierarchical Majorana
neutrinos.
 In the same
Figure (1), using dot-dashed lines, we have plotted the limits of
mixing angles obtained by assuming inverted hierarchy against the
lightest neutrino mass. It is immediately clear from Figures (1a)
and (1c) that the obtained ranges of $\theta_{12}$ and
$\theta_{23}$ are far from their experimental limits, thus ruling
out the inverted hierarchy for this particular combination.

Coming to the degenerate scenarios of Majorana neutrinos
characterized by either $m_{\nu_1} \lesssim m_{\nu_2} \sim
m_{\nu_3} \sim 0.1~\rm{eV}$ or $m_{\nu_3} \sim m_{\nu_1} \lesssim
m_{\nu_2} \sim 0.1~\rm{eV}$ corresponding to normal hierarchy and
inverted hierarchy respectively, one can easily infer
from Figures (1a) and (1c) that
degenerate scenario is also ruled out. This can be understood by noting that
around 0.1\rm{eV}
the limits obtained by assuming normal hierarchy and
inverted hierarchy have no overlap with experimental limits of
$\theta_{12}$ and $\theta_{23}$.

The viability of rest of the 143 combinations can similarly be checked
for normal hierarchy, inverted hierarchy and for degenerate Majorana
neutrinos.
For normal hierarchy, we find that in Category
(1), there are 12  viable combinations. Numerical results corresponding
to one such parallel combination, labelled as $ {\rm I}_a{\rm I}_a$, where both $M_l$ and $M_{\nu D}$ are of type  $ {\rm I}_a$, are given in
the first row of Table (\ref{at6}). The spectrum of neutrino masses shows that
neutrinos are following strict normal hierarchy. Further,
 $\theta_{12}$ and $\theta_{13}$ are spanning
their full experimental range, however the obtained range of
$\theta_{23}$ is just below its maximal value. One finds that
$\theta_{23}$ is sensitive to variations in the mass squared
differences, however it is not possible to obtain a higher value
for $\theta_{23}$ even when the ranges of $\Delta m_{12}^{2}$ and
$\Delta m_{23}^{2}$ are extended further.

 Apart from mixing
angles, we have also calculated the
 Jarlskog's rephasing invariant in the leptonic sector $J_{l}$ and 
  Dirac like CP violating phase
$\delta_{l}$ and effective neutrino mass $\langle m_{ee} \rangle $ related to
neutrinoless double beta decay $\beta \beta_{0 \nu}$.
 The parameter $J_{l}$ has been calculated by using
the expression
\be
J_l = Im[U_{23} U^{*}_{33} U_{22} U^{*}_{32}], \ee where $U_{23}$,
$U_{33}$, $U_{22}$ and $U_{32}$ are the elements of mixing matrix
U given in Equation (\ref{mixreal}).

The Dirac like CP violating phase $\delta_{l}$ can be determined
from
\be
J_l = s_{12}s_{23}s_{13}c_{12}c_{23}c^{2}_{13}{\rm sin} \delta_{l}
\ee where $s_{12}$, $s_{13}$ and $s_{23}$ are sines of mixing
angles $\theta_{12}$, $\theta_{13}$ and $\theta_{23}$.

The effective Majorana mass to be measured in $\beta \beta_{0 \nu}$
decay experiment, is given as
\be
\langle m_{ee} \rangle =|{m_{\nu_1}} U^{2}_{11} + {m_{\nu_2}} U^{2}_{12} +
 {m_{\nu_3}} U^{2}_{13}|.
\ee
The obtained ranges of $J_{l}$, $\delta_{l}$ and $\langle m_{ee} \rangle $ 
 are in agreement with the ranges
obtained in other such analyses.\cite{nonflavor,garcia} \  It may
be mentioned that the results corresponding to other parallel
combinations e.g. ${\rm I}_b{\rm I}_b$, ${\rm I}_c{\rm I}_c$,
 ${\rm I}_d{\rm I}_d$, ${\rm I}_e{\rm I}_e$ and
 ${\rm I}_f{\rm I}_f$ are exactly same as $ {\rm I}_a{\rm I}_a$
and hence are not given in the table.

Similarly, Category (1) also has 6 non parallel viable
combinations such as
${\rm I}_a{\rm I}_b$, similar in predictions to combinations
${\rm I}_b{\rm I}_a$,${\rm I}_c{\rm I}_f$,
${\rm I}_f{\rm I}_c$,
 ${\rm I}_d{\rm I}_e$, ${\rm I}_e{\rm I}_d$, given in the second row
of the Table (\ref{at6}). It is clear from the table that the results for the non
 parallel combinations are similar
to the parallel combinations except that the obtained range of
$\theta_{23}$ is above its maximal value.
 Further, Categories (2) and (3) do not have
any viable combination, because of only two generation mixing in
the neutrino mass matrix. Lastly, Category (4) does have four
viable combinations such as ${\rm II}_f{\rm I}_a$  similar in predictions 
 to
  ${\rm II}_d{\rm I}_a$; and
 ${\rm II}_d{\rm I}_b$, 
similar in predictions to ${\rm II}_f{\rm I}_b$,  given
respectively in the third and fourth row of the Table (\ref{at6}). The above two
sets can be distinguished again on the basis of $\theta_{23}$, as
combination  ${\rm II}_f{\rm I}_a$ gives $\theta_{23}$ below the
maximal value while
 ${\rm II}_d{\rm I}_b$ gives $\theta_{23}$ above its maximal value.
 All the four
combinations of Category (4) lead to a constrained range of $\theta_{13}$ e.g.;
$7^\circ -9^\circ$, very much in compliance with the latest data.
 It may be noted that although most of the
phenomenological implications of the above mentioned 16 texture six
zero lepton mass matrices are similar, however still these matrices can be
experimentally distinguished with more precise measurements of
$\theta_{23}$ and $\theta_{13}$.
 The above
mentioned texture combinations are found to be compatible with the
current data even when the inputs are at their 2$\sigma$ C.L..

Similarly, the viability of texture six zero hermitian 
lepton mass matrices can be checked for inversely hierarchical
Majorana neutrinos. Our analysis shows that inverted hierarchy as well as
degenerate neutrinos are completely ruled out for hermitian texture
six zero lepton mass matrices.

\begin{table}
{\setlength\arraycolsep{-0.5pt} 
\caption {Calculated ranges of neutrino masses, mixing angles,
$\bf \langle m_{ee} \bf\rangle $, $J_{l}$ and $\delta_{l}$
for viable combinations of $M_{l}$ and $M_{\nu}$ for normally
hierarchical Majorana neutrinos.}\label{at6}
\bt{ccllccc} \hline\hline & & & & & &
\\
Category& $\rm \bf M_l$ ~~~ $\rm \bf M_{\nu D}$ &  Neutrino
masses & Mixing angles & $\bf \langle m_{ee} \bf\rangle $ &
 $  \bf J_{l}$ & $ \bf \delta_{l}$ \\ & & & & & & \\
  \hline
& & & & & & \\ & ${\rm I}_{a}~~~~~~~~~{\rm I}_{a}$
 & & & & & \\
 1 & \scriptsize{$\left (\ba{ccc} 0 & A_l & 0    \\
A^{*}_l & 0 & B_l     \\
 0 & B^{*}_l   & C \ea \right)$}\scriptsize {$\left ( \ba{ccc}
0 & A_{\nu} & 0  \\ A^{*}_{\nu} & 0 & B_{\nu}     \\
 0 & B^{*}_\nu   & C \ea \right)$} &
\scriptsize {$\ba{c} m_{\nu_1}=0.0007-0.0032 \\
m_{\nu_2}=0.0084-0.0093 \\ m_{\nu_3}=0.0486-0.0516 \ea $} &
\scriptsize {$\ba{c} \theta_{12}= 31^o-35^o \\ \theta_{13}=
7.7^o-10.4^o\\ \theta_{23}=35^o-43^o \ea $} &
 \scriptsize {0.0029-0.0081}&
 \scriptsize {0.0-0.014} &
\scriptsize {$0-20^o$} \\ & & & & & & \\
 \hline
& & & & & &\\

& ${\rm I}_{a}~~~~~~~~~{\rm I}_{b}$
 & & & & & \\
 1 & \scriptsize  {$\left( \ba{ccc} 0 & A_l & 0    \\
A^{*}_l & 0 & B_l     \\
 0 & B^{*}_l   & C_l \ea \right)$}\scriptsize  {$\left( \ba{ccc}
0 & 0 & A_{\nu}      \\ 0 & C_{\nu} & B_{\nu}     \\ A^{*}_{\nu} &
B^{*}_\nu   & 0 \ea \right)$} & \scriptsize {$\ba{c}
m_{\nu_1}=0.0007-0.0033 \\ m_{\nu_2}=0.0083-0.0098 \\ m_{\nu_3} =
0.0486 - 0.0517 \ea $}
 &\scriptsize  {$\ba{c} \theta_{12}= 31^o-38^o \\
\theta_{13}= 8^o-11^o\\ \theta_{23}=44^o-54^o \ea $}    &
\scriptsize {0.0034-0.0065} & \scriptsize {0.0-0.015} & \scriptsize
{$0-19^o$} \\ & & & & & &\\
 \hline

& & & & & & \\ & ${\rm II}_{f}~~~~~~~~~{\rm I}_{a}$
 & & & & & \\
4 & \scriptsize  {$\left( \ba{ccc} C_l & 0 & 0    \\ 0 & 0 & A_l
\\
 0 & A^{*}_l   & D_l \ea \right)$}\scriptsize  {$\left( \ba{ccc}
0 &  A_\nu & 0     \\ A^{*}_\nu & 0 & B_\nu     \\ 0 & B^{*}_\nu &
C_\nu \ea \right)$} & \scriptsize {$\ba{c} m_{\nu_1}=0.0295-0.0442
\\ m_{\nu_2}=0.0918-0.0949 \\ m_{\nu_3}=0.220-0.227 \ea$}
 & \scriptsize{$\ba{c} \theta_{12}= 31^o-37^o \\
\theta_{13}= 7.7^o-8.6^o\\ \theta_{23}=35^o-43^o \ea $}  &
\scriptsize {0.0037-0.0061}&
 \scriptsize { 0.0-0.027} &
\scriptsize { $0-55^o$}  \\ & & & & & &\\
 \hline
& & & & & &\\ & ${\rm II}_{d}~~~~~~~~~{\rm I}_{b}$
 & & & & & \\

4 & \scriptsize {$\left( \ba{ccc} C_l & 0 & 0    \\ 0 & D_l & A_l
\\
 0 & A^{*}_l   & 0 \ea \right)$} \scriptsize  {$\left( \ba{ccc}
0 & 0 & A_\nu      \\ 0 & C_\nu & B_\nu     \\ A^{*}_\nu &
B^{*}_\nu   & 0 \ea \right)$} & \scriptsize {$\ba{c}
m_{\nu_1}=0.0291-0.0501 \\ m_{\nu_2}=0.0914 - 0.0969 \\
m_{\nu_3}=0.2155-0.2324 \ea$}
 &\scriptsize {$\ba{c} \theta_{12}= 31^o-38^o \\
\theta_{13}= 7^o-9^o\\ \theta_{23}=46^o-55^o \ea $} & \scriptsize
{0.0037-0.0061}&
 \scriptsize {0.0-0.011} &
\scriptsize {$0-20^o$}  \\ & & & & & & \\
 \hline
\et }
\end{table}

\section{DIRAC NEUTRINOS}

Coming to the case of Dirac neutrinos, we have again analyzed
 144 combinations for
normal hierarchy, inverted hierarchy and degenerate neutrinos.
 For comparison
with the Majorana neutrino case, we pick up
 the same non-Fritzsch like combination given
in Equation (\ref{nfe}), to check its compatibility
 with the latest neutrino mixing data. The diagonalizing transformations
for these matrices can easily be obtained in terms of neutrino
masses $m_{\nu_1}$, $m_{\nu_2}$ and $m_{\nu_3}$ for both normal and
inverted hierarchy  and are given in the Appendix (A). The
phase matrix $ P_{\nu D}$ and the scanned ranges of lightest neutrino mass
and phases $\phi_1$,  $\phi_2$and $\phi_3$ have already been mentioned above.

\begin{figure}[hbt]
\centerline{\includegraphics[width=10cm,height=7cm]{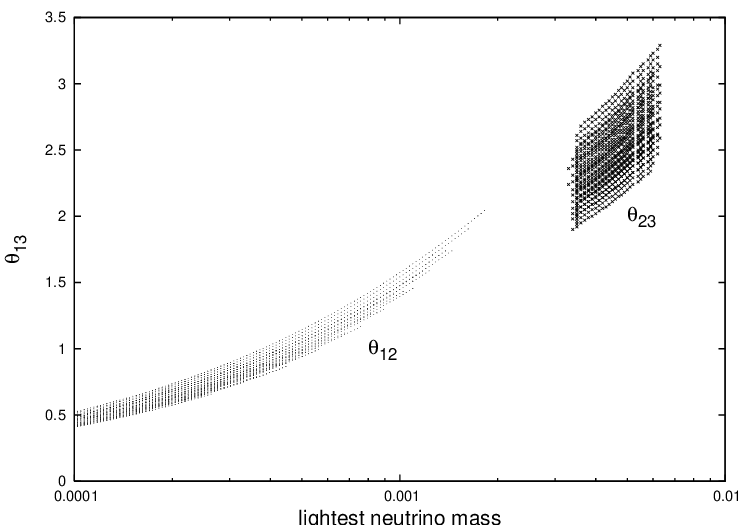}}
 \caption{Allowed parameter space of $\theta_{12}$
and $\theta_{23}$ in `lightest neutrino mass - $\theta_{13}$
plane' for normal hierarchy of Dirac neutrinos for texture six
zero combination given in Equation (\ref{nfe}). The dots and
crosses represent the allowed parameter space for $\theta_{12}$
and $\theta_{23}$ respectively.} \label{dirac}
\end{figure}

To check the compatibility of this particular combination for
normal hierarchy of Dirac neutrinos,
in Figure (\ref{dirac}) we have plotted the allowed parameter space for
$\theta_{12}$ and $\theta_{23}$ in
$m_{\nu_1}$-$\theta_{13}$ plane,
 represented respectively by dots and crosses.
 A general look at the figure shows that
there is no common parameter space available to
$\theta_{12}$ and $\theta_{23}$. Moreover, the obtained  $\theta_{13}$
range is much below the experimental range,
concluding that this particular combination
is not viable  for Dirac
neutrinos. This result remains unaffected even if the input parameter
ranges are extended further.
 Thus, normally
 hierarchical Dirac neutrinos are ruled out for the texture
combination given in Equation (\ref{nfe}).
Similarly, one can also show that degenerate and inversely
hierarchical Dirac neutrinos for this particular combination are
also ruled out.

The combinations which are viable for Majorana neutrinos, are
not viable for Dirac neutrinos primarily because of mixing angle $\theta_{23}$.
For example, for parallel combinations given in the first row
as well as the combinations given in the third row of the Table (\ref{at6}),
$\theta_{23}$ for dirac neutrinos comes out to be below the experimental limits.
Similarly, for non parallel combinations given in the second row and
the combinations given in the fourth row of the Table (\ref{at6}),
 $\theta_{23}$ lies above its experimental limits.

A similar analysis carried out
 for the rest of the 128 combinations shows that
there are no viable texture six zero lepton mass matrices for normally hierarchical,
inversely hierarchical as well as for degenerate Dirac neutrinos,
thus ruling out Dirac neutrinos completely for texture six zero mass matrices.

For the sake of completion, we have also analysed the cases
 corresponding to charged leptons being in the flavor basis
for  Dirac as well as
Majorana neutrinos and one finds that
 none of matrices gives result within the experimental ranges.

\section{SUMMARY AND CONCLUSION}
To summarize, we have analyzed 144 combinations corresponding to 
hermitian texture six zero lepton mass
matrices to check for their viability against current neutrino oscillation
data. For each combination of mass matrices, various cases
have been considered in the analysis, for example, normal
hierarchy, inverted hierarchy and degenerate neutrinos for both
Majorana as well as Dirac neutrinos. For Majorana neutrinos with normal
hierarchy out of 144, only 16 combinations are compatible with
current neutrino oscillation data at $3\sigma$ C.L..
  The above mentioned texture combinations are found to
be compatible with current data even at $2\sigma$ level. The 16 viable combinations can be grouped in four sets such that the matrices placed in each set are 
similar with regards to their predictions for lepton masses and flavor mixing parameters.  It is important to note that these four sets can be experimentally distinguished from each other with more precise measurements of $\theta_{13}$ and  $\theta_{23}$.
The
ranges of neutrino masses, the PMNS matrix, Jarlskog's rephasing
parameter $J_{l}$, Dirac like CP violating phase $\delta_{l}$
 and effective neutrino mass $\langle m_{ee}
\rangle $, calculated for each of the viable combination, are in
agreement with the ranges obtained in other such analyses. We find that 
inverted hierarchy and degenerate neutrinos are ruled out for
texture six zero Majorana neutrinos. Interestingly, for Dirac
neutrinos none of the 144 combinations is viable for normal
hierarchy, inverted hierarchy as well as degenerate neutrinos,
thus ruling out Dirac neutrinos for texture six zero lepton mass
matrices.

\section*{Acknowledgements} 

N.M. and P.S. Gill would like to thank the Principal, DAV College  and the Principal, SGGS College, respectively, for providing facilities to work.
M.R. is supported by the UGC, Govt. of India, under the Research Award Scheme (No.F.30-39/2011(SA-II)).


\appendix 
\section{The diagonalization transformations}
The diagonalization transformations
for the real texture three zero mass
matrices given in Equation (\ref{nfe}) are as follows:
\begin{enumerate}
\item   The diagonalizing matrix $O_l$ for the real texture three zero mass matrix
  $M_l$,
 \be
 M_{l}=\left( \ba{ccc}
C_{l} & 0 & 0      \\ 0 & D_{l} &  A_{l}     \\
 0 &  A_{l}^{*}  &  0 \ea \right),
\ee
is given as \\
\be
O_l = \left( \ba{ccc}
1 & 0 & 0      \\ 0 & \sqrt\frac{m_{\mu}}{m_{\mu} + m_{\tau}} &
 \sqrt\frac{m_{\tau}}{m_{\mu} + m_{\tau}}     \\
 0 & - \sqrt\frac{m_{\tau}}{m_{\mu} + m_{\tau}}
 & \sqrt\frac{m_{\mu}}{m_{\mu} + m_{\tau}}  \ea \right),
\ee
where $m_{\mu}$ and $m_{\tau}$ are the masses of the charged leptons $\mu$ and $\tau$.
\item
The diagonalizing matrix $O_{\nu D}$ for the real texture three zero neutrino mass
matrix $M_{\nu D}$,
\be
 M_{\nu D}=\left( \ba{ccc}
 0 & 0 & A_{\nu D}      \\ 0 & C_{\nu D} &  B_{\nu D}     \\
 A_{\nu D}^{*}  & B_{\nu D}^{*} & 0 \ea \right),
\ee
for normal hierarchy of neutrinos is given as  \\
\\
{\scriptsize $O_{\nu D}$ =
 $ \left( \ba{ccc}
-\sqrt\frac{m_{2}m_{3}(m_{3}-m_{2})}{(m_{1} - m_{2} + m_{3})
(m_{1}+m_{2})(m_{3}-m_{1})}  &
 \sqrt\frac{m_{1}m_{3}(m_{1}+m_{3})}{(m_{1} - m_{2} + m_{3})
(m_{1}+m_{2})(m_{3}+m_{2})} &
 \sqrt\frac{m_{2}m_{1}(m_{2}-m_{1})}{(m_{1} - m_{2} + m_{3})
(m_{3}+m_{2})(m_{3}-m_{1})}     \\
 \sqrt\frac{m_{1}(m_{1}+m_{3})(m_{2}-m_{1})}{(m_{1} - m_{2} + m_{3})
(m_{1}+m_{2})(m_{3}-m_{1})} &
 \sqrt\frac{m_{2}(m_{2}-m_{1})(m_{3}-m_{2})}{(m_{1} - m_{2} + m_{3})
(m_{1}+m_{2})(m_{2}+m_{3})} &
 \sqrt\frac{m_{3}(m_{1}+m_{3})(m_{3}-m_{2})}{(m_{1} - m_{2} + m_{3})
(m_{3}+m_{2})(m_{3}-m_{1})}  \\
 \sqrt\frac{m_{1}(m_{3}-m_{2})}{(m_{1}+m_{2})(m_{3}-m_{1})} &
 \sqrt\frac{m_{2}(m_{1}+m_{3})}{(m_{1}+m_{2})(m_{3}+m_{2})} &
 \sqrt\frac{m_{3}(m_{2}-m_{1})}{(m_{3}-m_{1})(m_{3}+m_{2})}
 \ea \right) $}.

\vspace{0.3cm}
Similarly, for inverted hierarchy of neutrinos, $O_{\nu D}$ is given as\\
\\
{\scriptsize
 $O_{\nu D}$
 = $\left( {\setlength\arraycolsep{0pt}  \ba{ccc}
 \sqrt\frac{m_{2}m_{3}(m_{3}+m_{2})}{(-m_{1}+m_{2} + m_{3})
(m_{1}+m_{2})(m_{3}+m_{1})}&
-\sqrt\frac{m_{1}m_{3}(m_{1}-m_{3})}{(-m_{1}+ m_{2} + m_{3})
(m_{1}+m_{2})(m_{3}+m_{2})}&
 \sqrt\frac{m_{2}m_{1}(m_{1}-m_{2})}{(-m_{1}+ m_{2} + m_{3})
(m_{3}+m_{1})(m_{3}-m_{2})} \\
-\sqrt\frac{m_{1}(m_{1}-m_{3})(m_{2}-m_{1})}{(-m_{1} + m_{2} + m_{3})
(m_{1}+m_{2})(m_{3}-m_{1})} &
 \sqrt\frac{m_{2}(m_{2}-m_{1})(m_{3}+ m_{2})}{(-m_{1}+m_{2} + m_{3})
(m_{1}+m_{2})(m_{2}-m_{3})} &
-\sqrt\frac{m_{3}(m_{2}+m_{3})(m_{3}-m_{1})}{(-m_{1} + m_{2} + m_{3})
(m_{3}- m_{2})(m_{3}+m_{1})} \\
 \sqrt\frac{m_{1}(m_{3}+m_{2})}{(m_{1}+m_{2})(m_{3}+m_{1})} &
 \sqrt\frac{m_{2}(m_{1}-m_{3})}{(m_{1}+m_{2})(m_{2}-m_{3})} &
 \sqrt\frac{m_{3}(m_{1}-m_{2})}{(m_{3}-m_{2})(m_{3}+m_{1})}
 \ea } \right)$},\\
\\
 where  $m_{1} = m_{\nu_1}$, $m_{2} = m_{\nu_2}$ and $m_{3} =
 m_{\nu_3}$ for Dirac neutrinos and
$m_{1} = \sqrt{m_{\nu_1} m_R}$,
$m_{2} = \sqrt{m_{\nu_2} m_R}$ and $m_{3} =\sqrt{m_{\nu_3} m_R}$
 for Majorana neutrinos.

\end{enumerate}
 
\section{$M_R$ with three texture zeros}
In this Appendix, we present some details of the texture six zero mass matrices, wherein along with $M_l$  and $M_{\nu D}$, we impose a  texture three zero structure 
on $M_R$ also.  $M_l$ and $M_{\nu D}$ can be any of the 12 matrices given in Table (\ref{t12}), while we choose $M_R$ to be similar in texture  structure to $M_{\nu D}$, thereby yielding 144 combinations of $M_l$ and $M_\nu$ which must be confronted against the neutrino oscillation data  given in Equations (\ref{solatmmass}) and (\ref{s13}). 
These 144 combinations can be divided in four categories.\vspace{0.2cm}

\noindent Category 1: $M_{l}$ from Class I, $M_{\nu D}$ and $M_R$ from Class I. \\
Category 2: $M_{l}$ from Class II, $M_{\nu D}$ and $M_R$ from Class II.\\
 Category 3: $M_{l}$ from Class I, $M_{\nu D}$ and $M_R$ from Class II. \\
 Category 4: $M_{l}$ from Class II, $M_{\nu D}$ and $M_R$  from Class I.\\

As an example, we choose a texture  combination ${\rm II}_b{\rm I}_e{\rm I}_e$ 
from Category (4),
where $M_l$ is of type ${\rm II}_b$ and  $M_{\nu D}$ and 
$M_R$ are of type ${\rm I}_e$, 
 as given below
\be
 M_{l}=\left( \ba{ccc}
0 & 0 &   A_{l}    \\ 0 & C_{l} & 0     \\
A_{l}^{*}  & 0 &  B_{l}\ea \right), ~~
M_{\nu D}=\left( \ba{ccc} 0 &
B_{\nu D} &  A_{\nu D}   \\ B_{\nu D}^{*} &  C_{\nu D}  &  0  \\
A_{\nu D}^{*} &  0  &  0 \ea \right),~~  M_R= 
\left( \ba{ccc} 0 & B_R &  A_R   \\ B_R &  C_R  &  0  \\
A_R &  0  &  0 \ea \right). \label{nmaj}
 \ee
For simplicity, we have neglected the  Majorana phases in the right-handed neutrino
mass matrix. 
Using seesaw Equation (\ref{seesaweq}), the effective neutrino mass matrix $M_{\nu}$ is given as
\be M_{\nu}=\left( \ba{ccc} D_{\nu} &
B_{\nu} &  A_{\nu}   \\ B_{\nu} &  C_{\nu}  &  0  \\
A_{\nu} &  0  &  0 \ea \right), \label{nmaj1} \ee 
 where 
\beqn A_{\nu} &= & |A_{\nu}|=-\frac{|A_{\nu D}|^2}{A_R} \\
B_{\nu} &= &-\frac{A_R B_{\nu D}^{*} C_{\nu D}-A_{\nu D}^{*} B_R C_{\nu D}+A_{\nu D}^{*}B_{\nu D} C_R}{A_R C_R}\\
C_{\nu} &=& |C_{\nu}| = -\frac{C_{\nu D}^2}{C_R} \\
D_{\nu} &= & -\frac{(A_R B_{\nu D}^{*} -A_{\nu D}^{*} B_R)^2}{A_R^2 C_R}  \eeqn
It is evident that $M_{\nu}$ turns out to be texture two zero type. However if the condition $A_R B_{\nu D}^{*} = A_{\nu D}^{*} B_R$ is satisfied, it becomes texture three zero i.e. the texture structure is preserved by the seesaw mechanism in that case. We find that such a simplified assumption does not yield viable results for matrices given in Equation (\ref{nmaj}).

 The matrices $M_l$ and  $M_{\nu}$ given above
can easily be diagonalized by bi-unitary transformations and the corresponding PMNS matrix can be constructed, as explained in Section (2).  
The  real diagonalizing matrix $O_l$ for $|M_l|$ 
 is given as,
\be 
 O_l
 = \left( \ba{ccc}
\sqrt\frac{m_{e}}{m_{e} + m_{\tau}}  & 0 & \sqrt\frac{m_{\tau}}{m_{e} + m_{\tau}}\\
 0 & 1 & 0    \\
 -\sqrt\frac{m_{\tau}}{m_{e} + m_{\tau}}& 0
 & \sqrt\frac{m_{e}}{m_{e} + m_{\tau}}
\ea \right),
\ee
where $m_e$ and $m_{\tau}$ are the masses of the charged leptons $e$ and $\tau$.
Similarly, the  real diagonalizing matrix  $O_{\nu}$ for  $|M_{\nu}|$ is given as,
\be 
 O_{\nu}
 = \left( \ba{ccc}\sqrt{\frac{m_{\nu_1} (C_{\nu}-m_{\nu_1})}{(m_{\nu_1}+m_{\nu_2})
   (m_{\nu_3}-m_{\nu_1})}} &
\sqrt{\frac{m_{\nu_2} (C_{\nu}+m_{\nu_2})}{(m_{\nu_1}+m_{\nu_2})
   (m_{\nu_2}+m_{\nu_3})}} &
\sqrt{\frac{m_{\nu_3} (m_{\nu_3}-C_{\nu})}{(m_{\nu_3}-m_{\nu_1})
   (m_{\nu_2}+m_{\nu_3})}}  
  \\
-\sqrt{\frac{m_{\nu_1} (C_{\nu}+m_{\nu_2}) (m_{\nu_3}-C_{\nu})}{C_{\nu} (m_{\nu_1}+m_{\nu_2})
   (m_{\nu_3}-m_{\nu_1})}}   &
 -\sqrt{\frac{m_{\nu_2} (C_{\nu}-m_{\nu_1}) (m_{\nu_3}-C_{\nu})}{C_{\nu} (m_{\nu_1}+m_{\nu_2})
   (m_{\nu_2}+m_{\nu_3})}}   &
\sqrt{\frac{m_{\nu_3} (C_{\nu}-m_{\nu_1}) (C_{\nu}+m_{\nu_2})}{C_{\nu} (m_{\nu_3}-m_{\nu_1})
   (m_{\nu_2}+m_{\nu_3})}} 
 \\
  \sqrt{\frac{m_{\nu_2} m_{\nu_3} (C_{\nu}-m_{\nu_1})}{C_{\nu} (m_{\nu_1}+m_{\nu_2})
   (m_{\nu_3}-m_{\nu_1})}}&
 -\sqrt{\frac{m_{\nu_1} m_{\nu_3} (C_{\nu}+m_{\nu_2})}{C_{\nu} (m_{\nu_1}+m_{\nu_2})
   (m_{\nu_2}+m_{\nu_3})}} &
 \sqrt{\frac{m_{\nu_1} m_{\nu_2} (m_{\nu_3}-C_{\nu})}{C_{\nu} (m_{\nu_3}-m_{\nu_1})
   (m_{\nu_2}+m_{\nu_3})}}
 \ea \right). \label{2zdiag}\ee
\noindent The   mixing matrix $U$ may be constructed by using the equation,
\be
 U = O_l^{\dagger} P_{l \nu} O_{\nu} \,, \label{mixmaj} \ee
where  $P_{l \nu}$ is the phase matrix, which in general can be taken as $(e^{i\phi_1},\,e^{i\phi_2},\,e^{i\phi_3})$;
$\phi_1$, $\phi_2$ and $\phi_3$ being
related to the phases of complex mass matrices $M_l$  and $M_{\nu}$.

Thus, in the mixing matrix, apart from the phases, 
$\phi_1$, $\phi_2$ and $\phi_3$, there comes an additional free parameter $C_{\nu}$.
As before, the phases have been given full variation from 0 to 2$\pi$, while we allow 
  $C_{\nu}$ to vary between 
$m_{\nu_1}$ and $m_{\nu_3}$, such that the diagonalizing transformation given in Equation 
(\ref{2zdiag}) remains real. Our results, 
with regards to mixing angles, CP violating phase
 $\delta_l$ and effective neutrino mass $\langle m_{ee} \rangle $,
given as, 
$\theta_{12} \approx 31^o-37^o,~ \theta_{13} \approx
7^o-9^o,~ \theta_{23}\approx44^o-45^o,~ \delta_{l}\approx 0-8^o,~
  \langle m_{ee}  \rangle \approx 0.0054 - 0.0074,$~
indicate that  the matrices given in Equation (\ref{nmaj}) accommodate the unsuppressed $\theta_{13}$ very well. The angle $\theta_{23}$ is maximal, however the CP violating phase
 $\delta_l$ is rather small for the combination  ${\rm II}_b {\rm I}_e {\rm I}_e$.

Similarly, in the Category (4), there are 11 more combinations  of $M_l$, $M_{\nu D}$ and 
$M_R$  which accommodate the data given in Equations (\ref{solatmmass}) and (\ref{s13}), for example  ${\rm II}_a {\rm I}_a {\rm I}_a$, ${\rm II}_a {\rm I}_b {\rm I}_b$, 
 ${\rm II}_b {\rm I}_d {\rm I}_d$, ${\rm II}_f {\rm I}_a {\rm I}_a$, ${\rm II}_f {\rm I}_b {\rm I}_b$,
 ${\rm II}_c {\rm I}_c {\rm I}_c$, ${\rm II}_c {\rm I}_f {\rm I}_f$, ${\rm II}_d {\rm I}_a {\rm I}_a$, ${\rm II}_d {\rm I}_b {\rm I}_b$,
 ${\rm II}_e {\rm I}_a {\rm I}_a$, ${\rm II}_e {\rm I}_b {\rm I}_b$. Most of these combinations have similar predictions with regards to $\theta_{12}$ and $\theta_{13}$, however the allowed  ranges  of $\theta_{23}$ and $\delta_{l}$ vary.

One of the  purpose of the above analysis is to compare with the case of diagonal $M_R$ discussed in detail in Section (5), we find that with texture three zero $M_R$, apart from the combinations given in Table (\ref{at6}), several new textures, as given above, also become viable.
 Unlike the diagonal $M_R$ case, we do not get any viable combination from Category (1), while similar to diagonal $M_R$ case, Categories (2) and (3) do not accommodate experimental data. To summarise the Appendix, we can say that the set of three zero mass matrices for $M_l$, $M_{\nu D}$ and $M_R$,  which can explain the oscillation data, is not unique. More refined measurements of mixing angles, particularly $\theta_{23}$ and also a measurement of CP violation in the leptonic sector can help isolate the unique texture structure for charged leptons and neutrinos.

\end{document}